# Mutation activity of *Lonicera caerulea* population in an active fault zone (the Altai Mountains)

I.G. Boyarskikh[1], A.I. Kulikova[1], A.R. Agatova[2,3], A.I. Bakiyanov[4], I.V. Florinsky[5*]

[1] Central Siberian Botanical Garden, Siberian Branch, Russian Academy of Sciences
101, Zolotodolinskaya St., Novosibirsk, 630090, Russia

[2] V.S. Sobolev Institute of Geology and Mineralogy, Siberian Branch, Russian Academy of Sciences
3, Academician Koptyug Ave., Novosibirsk, 630090, Russia

[3] Ural Federal University
19, Mira St., Ekaterinburg, 620002, Russia

[4] Gorno-Altaisk State University
1, Lenkina St., Gorno-Altaisk, Altai Republic, 649000, Russia

[5] Institute of Mathematical Problems of Biology, Russian Academy of Sciences
Pushchino, Moscow Region, 142290, Russia

## Abstract

Geophysical and geochemical anomalies may have a mutagenic effect on plants growing in active fault zones being the factors of evolutionary transformation of plant populations. To test this assumption we evaluated the mutation activity of a *Lonicera caerulea* natural population in one of the active fault zones in the Altai Mountains. We derived principal cytogenetic indices (i.e., mitotic, prophase, metaphase, anaphase, and telophase indices as well as proportion and range of abnormal mitoses) for meristematic cells of *Lonicera caerulea* seedlings. We found that the local geological and geophysical environment (i.e., mineralogical composition of rocks and anomalies of the magnetic field) increases the mitotic activity and the number of abnormal mitoses in the meristematic cells. The results may help to clarify the role of environmental conditions of tectonically active regions in microevolutionary processes.

## 1. Introduction

Active faults are channels for matter and energy coming from the depths to the surface and influencing the environment. Upward fluid migration and gas emanation leads to the formation of geochemical anomalies in active fault zones. Gravity and magnetic anomalies are also often associated with these zones (Handy et al., 2007). Geophysical and geochemical anomalies of active fault zones can cause a wide range of responses in biota (Florinsky, 2010). Active faults can control the spatial distribution of vegetation and development of azonal biocenoses (Vinogradov, 1955; Bgatov et al., 2007), impact the chemical composition of plants, and increase variability of plant morphological, phenotypic, and biochemical attributes (Trifonov, Karakhanian, 2004; Kutinov et al., 2009; Boyarskikh, Shitov, 2010; Vyukhina et al., 2013). Large lateral movements along faults affect the evolution, distribution, and taxonomic diversity of biocenoses (Heads, 1994, 1998, 2008). It is well known that hotspots of biodiversity – for example, most centers of the greatest diversity of cultivated plants (Vavilov, 1940) – are located in topographically complex mountain regions. This is recently interpreted as an indirect result of the tectonically driven orogenesis leading to changes in topography that, in turn, alter (micro)climatic conditions, population areas, and connectivity between populations (Badgley, 2010; Kent-Corson et al., 2013; Favre et al., 2015). However, the fact that biodiversity hotspots correlate with tectonically active areas gave rise to the hypothesis that geophysical and geochemical anomalies of active faults may also influence speciation processes (Syvorotkin, 2010).

[*] Correspondence to: iflor@mail.ru

An increase in the number of chromosomal aberrations in cells of meristematic tissues is a nonspecific reaction of cells to the action of environmental mutagens. Cytogenetic analysis (Grant, 1978, 1998) can be used to evaluate the mutation activity in plant populations within active fault zones. Perennial higher plants are suitable for such a test since they can detect an effect of prolonged exposure. We previously chose *Lonicera caerulea* L. family *Caprifoliaceae* Juss. (blue honeysuckle) as a model object for our research in the Altai Mountains (Boyarskikh, Shitov, 2010; Boyarskikh et al., 2012a, 2012b; Kulikova, Boyarskikh, 2014) because this species is distributed throughout mountain regions of Central Eurasia dominating in forest shrubs.

We earlier described a population of *Lonicera caerulea* subsp. *altaica* Pall. near the Village of Verkh-Uimon (the Ust-Koksa Region, Altai Republic, Russia) (Boyarskikh et al., 2012b). In this population, we observed a sharp increase in the polymorphism of morphological traits of blossoms and changes in their functional status. There were plants with various types of fasciated blossoms, untypical locations of androecium and gynoecium, abnormalities in the structure of the anthers and pollen grains, abnormalities in microsporogenesis, and cytomixis (Kulikova, Boyarskikh, 2014). In this paper, we analyze features of mitosis in seedling cells from this population to estimate mutagenic effects within an active fault zone and their possible role in microevolutionary transformation of populations. We study a cytogenetic response on the action of an undifferentiated set of factors of the geological environment rather than its individual factors.

## 2. Study area

We conducted the study at the northwestern margin of the Katun Range, the high-mountain part of Altai (Fig. 1A). The region is seismically active in the Late Pleistocene–Holocene (Lukina, 1996; Deev et al., 2013). The study area is situated on the Molnieboi Spur at an altitude of 1,250–1,300 m above sea level. The Molnieboi Spur has a length of about 2.4 km, a width of up to 100 m, and a relative height of up to 500 m above the adjacent Okol depression-graben (Fig. 1B).

The Molnieboi Spur has a stepped shape caused by a system of reactivated ancient and modern local faults (Fig. 1B). Zones of these faults are topographically manifested as scarps on the crest of the spur, and gullies on its slopes. Slickensides of serpentine testify motions along these faults. The current activity of these faults is indicated by repeating earthquake-driven radon emissions recorded during our studies. They are apparently associated with the active sublatitudinal Kucherla-Uimon fault (Fig. 1B) situated about 3 km northward from the Molnieboi Spur (Deev et al., 2013).

The Molnieboi Spur is composed from a rock plate of the epidote-amphibolite facies of metamorphism including alternating green (quartz-chlorite) schists, amphibolites, granites, and granite-gneisses (Boyarskikh et al., 2012b) (Table 1). There are solitary plagioclase porphyry dykes and skarn zones. Green schists have undergone metasomatic alteration and include ore minerals, such as hematite and magnetite. All rocks are highly fragmented and fractured. The visible thickness of the plate exceeds 1,500 m. Differences in the mineralogical composition of the rocks determine the heterogeneity in the spatial distribution of chemical elements and radionuclides in the soil (Table 2).

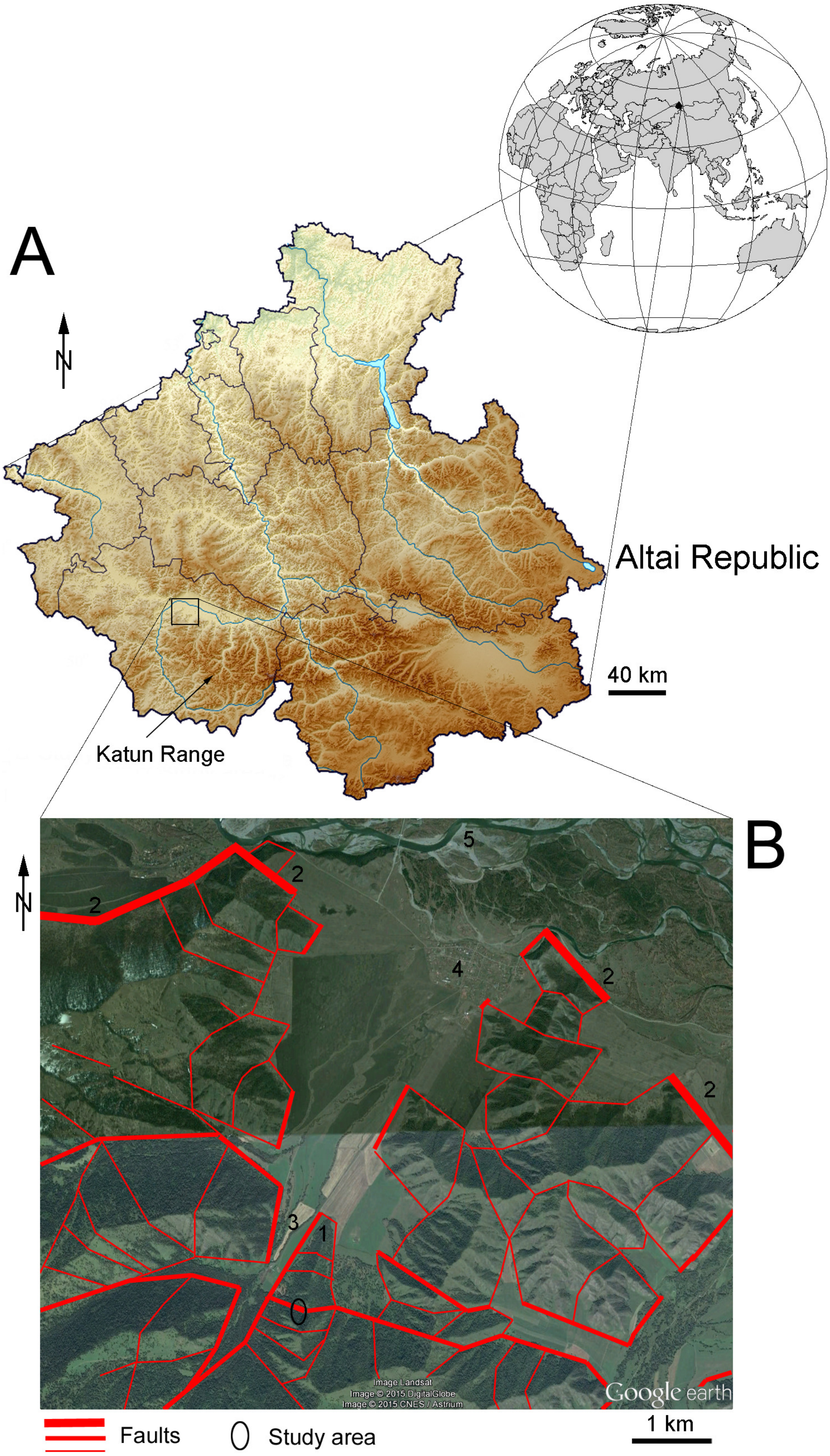

Fig 1. Location of the study area (50º10'30" N, 85º42'30" E). (A) Regional scale (the Altai Mountains). (B) Local scale (the network of local faults on a remotely sensed image mosaic): the Molnieboi Spur (1), the Kucherla-Uimon fault (2), the Okol depression-graben (3), the Village of Verkh-Uimon (4), the Katun River (5).

Table 1. Geographic, petrographic, and botanical characteristics of the study area.

| Site | Landscape position, average elevation above sea level | Rocks | Vegetation |
|---|---|---|---|
| C | Western slope, 1,297 m | Quartz-chlorite schist with interbedded hematite and epidote veinlets. | The larch and birch shrub forest: *Larix sibirica*, *Betula pendula*.<br>The medium-density undergrowth: *Cotoneaster melanocarpus*, *Lonicera caerulea* subsp. *altaica*, *Pinus sibirica, Populus tremula*, *Ribes atropurpureum, Rosa pimpinellifolia, Rosa acicularis, Rubus idaeus, Rubus saxatilis, Spiraea chamaedryfolia.*<br>The patchy herbaceous cover: *Aconitum septentrionale, Aegopodium alpestre, Atragene sibirica, Bupleurum longifolium, Cimicifuga foetida, Cacalia hastata, Calamagrostis obtusata, Cerastium pauciflorum, Chamaenerion angustifolium, Cruciata krylovii, Dactylis glomerata, Elymus mutabilis, Galium boreale, Lathyrus frolovii, Paeonia anomala, Pleurospermum uralense, Poa sibirica, Pulmonaria mollis, Thalictrum minus, Trollius asiaticus, Vicia cracca, Vicia sepium.* |
| A+ | Western slope, 1,291 m | Quartz-chlorite schist with plenty of magnetite and hematite; epidote, garnet, quartz, hematite, and magnetite in veinlets. | |
| A- | | Quartz-chlorite schist with hematite and epidote veinlets. | |
| W | Western upslope, 1,270 m | Granite, microgneisses, quartz-chlorite massive rock with a small amount of fine-grained hematite. | |
| E | Eastern upslope, 1,261 m | Quartz-chlorite schist, garnet-bearing skarn, granite. | |

Table 2. Total content of elements and radionuclide activity in soils of the study area (modified from Boyarskikh et al., 2012b).

| Site | Total content of elements, mg/kg | | | | | | | | Radionuclide activity, Bq/kg | | | | | |
|---|---|---|---|---|---|---|---|---|---|---|---|---|---|---|
| | K | Ca | Mn | Fe | Ni | Cu | Zn | Sr | $Ra^{226}$ | $Th^{232}$ | $K^{40}$ | $Cs^{137}$ | $Be^{7}$* | $Pa^{234}$ |
| C | 16,990 | 24,506 | 2,196 | 59,121 | 32 | 54 | 118 | 167 | 5.9 | 19 | 356 | 2 | 290 | <MDA |
| A+, A- | 14,171 | 33,462 | 2,012 | 78,631 | 20 | 61 | 123 | 184 | 2.4 | 25 | 320 | 7 | 840 | 300 |
| W | 23,949 | 17,934 | 1,294 | 49,152 | 20 | 23 | 72 | 139 | 24 | 47 | 860 | 5 | <MDA | <MDA |
| E | 22,295 | 35,847 | 2,796 | 66,701 | 35 | 37 | 150 | 192 | 21 | 22 | 749 | 5 | <MDA | <MDA |

MDA – minimum detectable activity; * $Be^{7}$ activity was calculated at the time of sampling.

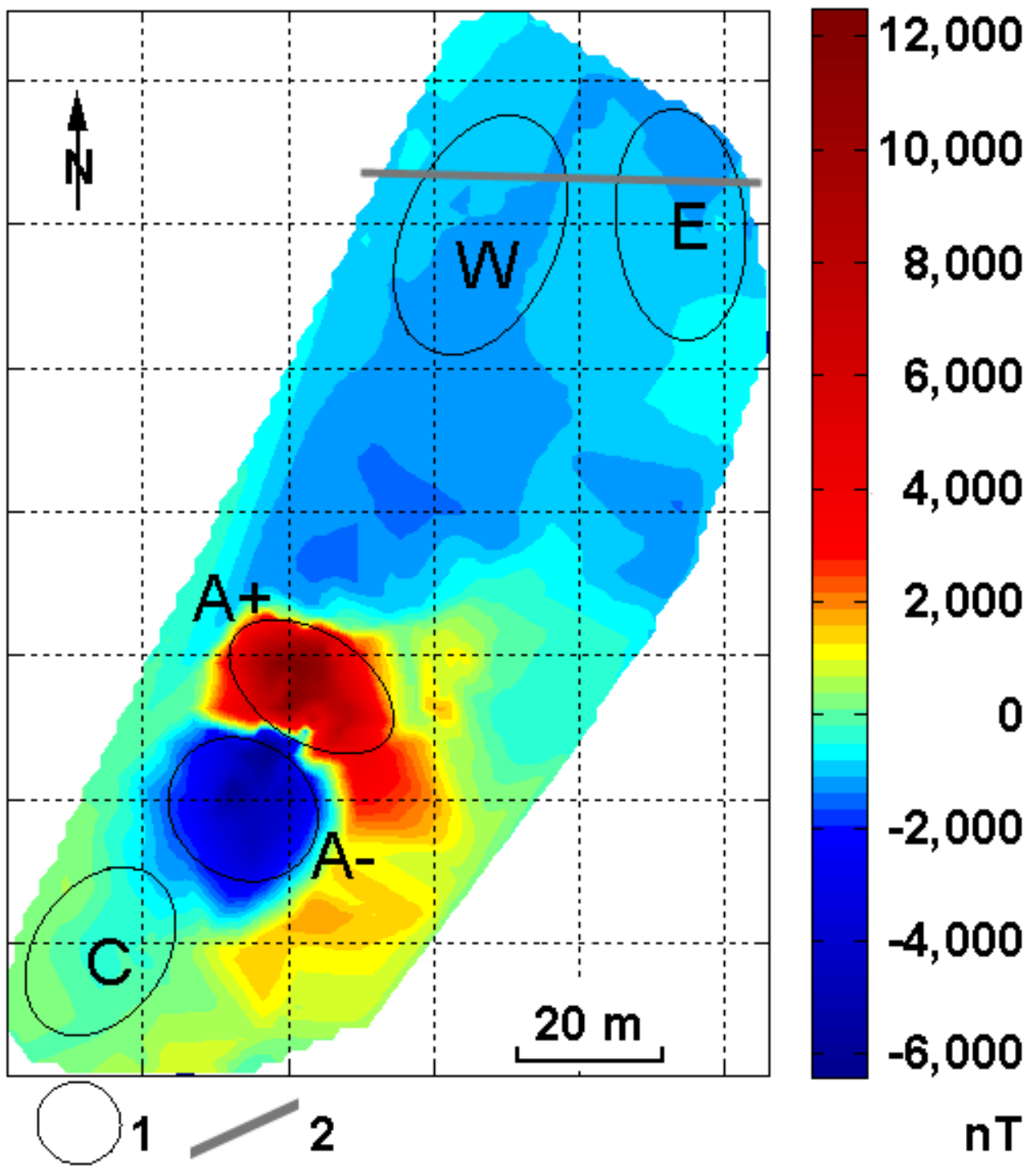

Fig. 2. Anomalous magnetic field within the study area. The normal magnetic field is 60,000 nT; sites (1), the local fault zone (2) (modified from Boyarskikh et al., 2012b).

Our earlier magnetometer measurements demonstrated very uneven spatial distribution of the magnetic field within the study area (Fig. 2). We revealed two pronounced anomalies, positive and negative, reminding "a dipole" (Boyarskikh et al., 2012b). In spite of the prolonged investigations of some parts of the Molnieboi Spur and adjacent areas (Dmitriev et al., 1989), the nature of the anomalies is not completely clear. We found that, in general, positive anomalies are spatially correlated with outcrops containing magnetite.

We previously selected five sites for detailed botanical, soil, and geological surveys (four test sites and one control site with sizes range from 300 to 600 $m^2$). The sites are located on the crest and adjacent upslopes of the upper part of the Molnieboi Spur, within an area with the size of about 150 × 100 m (Fig. 2). The sites are marked by similar temperature and water regimes as well as soil properties (granulometric composition, pH, organic matter content) (Boyarskikh et al., 2012b) that are indirectly confirmed by a similar plant species composition (Table 1). However, the sites differ in geological (Table 1), geochemical (Table 2), and geophysical characteristics, first of all, in the magnetic field (Fig. 2). At the altitude of about 1,290 m, there are the A+ site in the positive magnetic anomaly (72,000 nT) and the A- site in the negative magnetic anomaly (54,000 nT). South-west from these sites, there is the C control site marked by the normal magnetic field of this region (about 60,000 nT) at the altitude of about 1,297 m. North-east from the A+ and A- sites, there are the E and W sites along the eastern and western upslopes, correspondingly, at the altitude of 1,260–1,270 m. The E and W sites are situated in the local fault zone (Fig. 2) with the magnetic field reduced by 500–1,000 nT.

### 3. Materials and methods

Field work. We identified and labeled all plants of the *Lonicera caerulea* subsp. *altaica* population within each site. Then, we sampled twenty normally developed mature fruits from each of twenty plants within each of the five sites (400 fruits from each site).

Laboratory work. We carried out the cytogenetic analysis of meristematic tissue samples from seedlings using a technique (Singh, 2003) modified for *Lonicera caerulea* subsp. *altaica* as follows: Seeds were taken from fruits (30 seeds per each plant), placed into Petri dishes on a wet filter paper (an individual dish for each plant), and grown in an incubator at 25°C for about a month. Seedlings with a rootlet length of 0.4–0.5 cm were fixed in ethanol : acetic acid (3:1) at 11:00 UTC +06:00 (we observed a maximum division in root meristems at that time). Then, we transferred seedlings to 70% ethanol and stored at 5°C. The fixed samples were etched in a 4% ferric alum for 20 minutes, and stained with hematoxylin heating over the flame of a spirit lamp to accelerate the staining. The samples were then crushed in a drop of chloral hydrate for brightening. To study and photograph meristematic cells of the samples, we used an upright microscope Axioskop-40, a digital camera AxioCam MRc5, and morphometric software AxioVision 4.6 (Carl Zeiss MicroImaging GmbH). See Table 3 for the number of meristematic cells observed.

Cytogenetic analysis. We derived the following cytogenetic indices for each site:

1. The mitotic index, a measure of the intensity of cell division, or mitotic activity. We calculated the mitotic index as a percentage of dividing cells to the total number of observed cells (Urry et al., 2013).

2. The prophase, metaphase, anaphase, and telophase indices used to estimate variations in the duration of mitosis phases and the regularity of the cell cycle (Alov, 1972). We calculated each index as a percentage of cells in a related mitotic phase to the total number of observed cells.

3. Proportions of abnormal mitoses calculated as a percentage of cells with mitotic abnormalities observed in metaphase, anaphase, and telophase to the total number of dividing cells in a related mitotic phase. The total proportion of abnormal mitoses was also calculated. We did not consider prophase because it is difficult to observe mitotic abnormalities in this phase.

Statistical analysis. We compared cytogenetic indices of seedling meristems from the A+, A-, E and W test sites with those from the C control site using Pearson's $\chi^2$ test (Glantz, 2012). The null hypothesis assumed that there was no difference between seedling meristems from a test site and the control site (that is, no mutagenic effects of the active fault zone). If the $\chi^2$ values exceeded the critical ones, the null hypothesis was rejected and the difference between cytogenetic indices was considered statistically significant (testifying mutagenic effects of the active fault zone). The statistical analysis was carried out with Statistica and MS Excel software.

## 4. Results

We found that mitotic division was normal in most meristematic cells. The maximum number of dividing cells in various mitotic phases was observed in the seedling meristems from the W site located in the local fault zone (Table 3). The prophase index of seedling meristems from the W site was higher than that from the C site (Tables 3 and 4). The metaphase indices of seedling meristems from all the sites, except the A- site, were higher than that from the C site. There was no dependence of the anaphase index on the plant growing location. The telophase indices of seedling meristems from the A- and E sites were lower than that from the C site.

The analysis showed the dependence of the proportion of abnormal mitoses on the plant growing location (Fig. 3). We observed the statistically significant increase in the total proportion of abnormal mitoses for all the sites, except the W site (Tables 3 and 4). The largest proportion of abnormal mitoses was found for seedling meristems from the negative magnetic anomaly (the A- site).

We recorded the following types of abnormal mitoses in seedling cells: (1) lagging of chromosomes in prometaphase; (2) irregular grouping of chromosomes in metaphase; (3)

bridges, lagging, and ejections of chromosomes in anaphase; (4) lagging of chromosomes, broken bridges, and formation of several chromosome groups in telophase (Fig. 4). The range of abnormal mitoses depends on the fruit sampling location (Fig. 5). For the C control site, we observed almost the full range of abnormal mitoses, and abnormalities were almost evenly distributed over mitotic phases: 38% in metaphase, 31% in anaphase, and 31% in telophase. The negative magnetic anomaly (the A- site) had the narrowest range of abnormal mitoses: we observed only three types there, namely, lagging of chromosomes in prometaphase (71%), irregular grouping of chromosomes in metaphase (5%), and lagging of chromosomes in anaphase (24%). The positive magnetic anomaly (the A+ site) had the widest range of abnormal mitoses: lagging of chromosomes dominated in prometaphase (75%); chromosome lagging, ejection, and bridges were observed in anaphase (20%). For the W site, the largest

Table 3. Cytogenetic indices of seedling meristems depending on the fruit sampling location.

| Site | Total number of cells observed | Mitotic index, % | Prophase index, % | Metaphase index, % | Anaphase index, % | Telophase index, % | Total proportion of abnormal mitoses, % |
|---|---|---|---|---|---|---|---|
| C | 4,474 | 63.3 | 57.9 | 1.7 | 0.8 | 2.9 | 7.6 |
| A- | 1,543 | 64.2 | 59.7 | 2.1 | 0.6 | 1.8** | 19.2** |
| A+ | 2,686 | 61.1 | 55.0 | 2.4** | 1.0 | 2.7 | 13.8** |
| W | 3,398 | 71.4** | 64.8** | 2.3** | 1.0 | 3.3 | 9.8 |
| E | 1,284 | 67.8 | 61.7 | 2.7** | 1.3* | 2.1* | 13.9* |

Statistically significant differences between cytogenetic indices of seedling meristems from the test sites and control site for P ≤ 0.05 (*) and 0.01 (**).

Table 4. Values of $\chi^2$-criterion between cytogenetic indices of seedling meristems from the test sites and control site (P-values are in brackets).

| Cytogenetic indices | Test site: A- | A+ | W | E |
|---|---|---|---|---|
| Mitotic index | 0.15 (0.6970) | 1.26 (0.2613) | **16.63** (0.0000) | 2.86 (0.0910) |
| Prophase index | 0.64 (0.4241) | 2.48 (0.1150) | **13.74** (0.0002) | 2.37 (0.1236) |
| Metaphase index | 2.47 (0.1162) | **7.6** (0.0058) | **6.09** (0.0136) | **8.96** (0.0028) |
| Anaphase index | 1.61 (0.2040) | 0.9 (0.3429) | 0.79 (0.3756) | 3.46 (0.0628) |
| Telophase index | **8.14** (0.0043) | 0.51 (0.4758) | 1.4 (0.2361) | **4.13** (0.0422) |
| Proportion of abnormal mitoses in metaphase | **12.34** (0.0004) | **11.24** (0.0008) | **8.47** (0.0036) | **5.77** (0.0163) |
| Proportion of abnormal mitoses in anaphase | 3.18 (0.0748) | 0.15 (0.7022) | 1.12 (0.2901) | 0.13 (0.7174) |
| Proportion of abnormal mitoses in telophase | 1.97 (0.1600) | 3.1 (0.0781) | **4.65** (0.0311) | 0.26 (0.6083) |
| Total proportion of abnormal mitoses | **12.71** (0.0004) | **6.73** (0.0095) | 1.13 (0.2871) | **4.33** (0.0375) |

Bold values of $\chi^2$ exceed the critical values $\chi^2_{0.10} = 2.71$, $\chi^2_{0.05} = 3.84$, and $\chi^2_{0.01} = 6.63$ indicating statistically significant differences between cytogenetic indices of seedling meristems from the test sites and control site. P-values are in brackets.

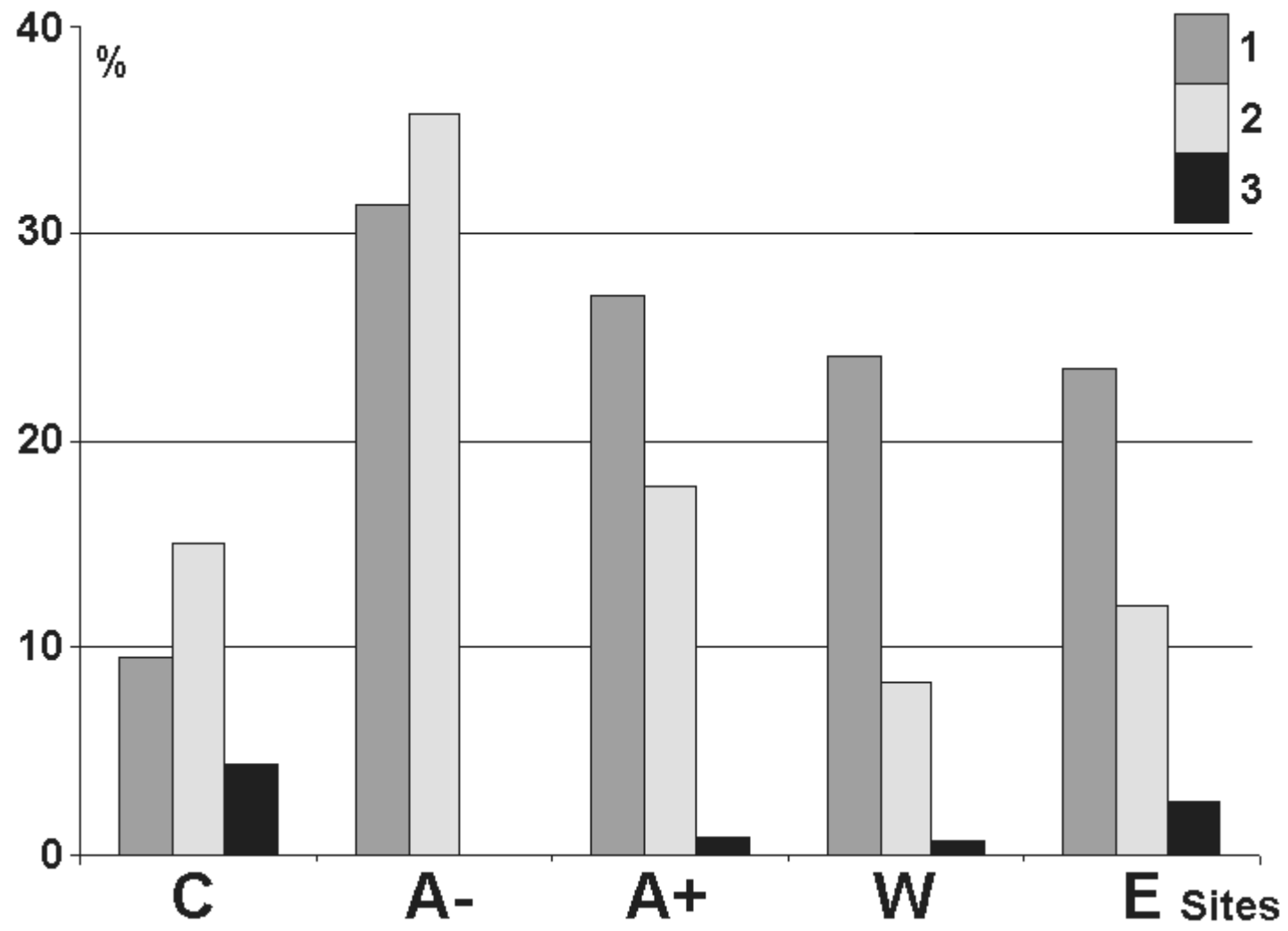

Fig. 3. Proportions of abnormal mitoses in meristematic cells of seedlings depending on the fruit sampling location. Abnormal mitoses in metaphase (1), anaphase (2), and telophase (3).

proportion of abnormal mitoses was found in metaphase (84%); there also were lagging of chromosomes and bridges in anaphase (13%). For the E site, we recorded the increase in mitotic abnormalities in metaphase: lagging of chromosomes in prometaphase (62%) and irregular grouping of chromosomes in metaphase (13%); smaller proportions of abnormal mitoses were observed in anaphase (19%) and telophase (6%).

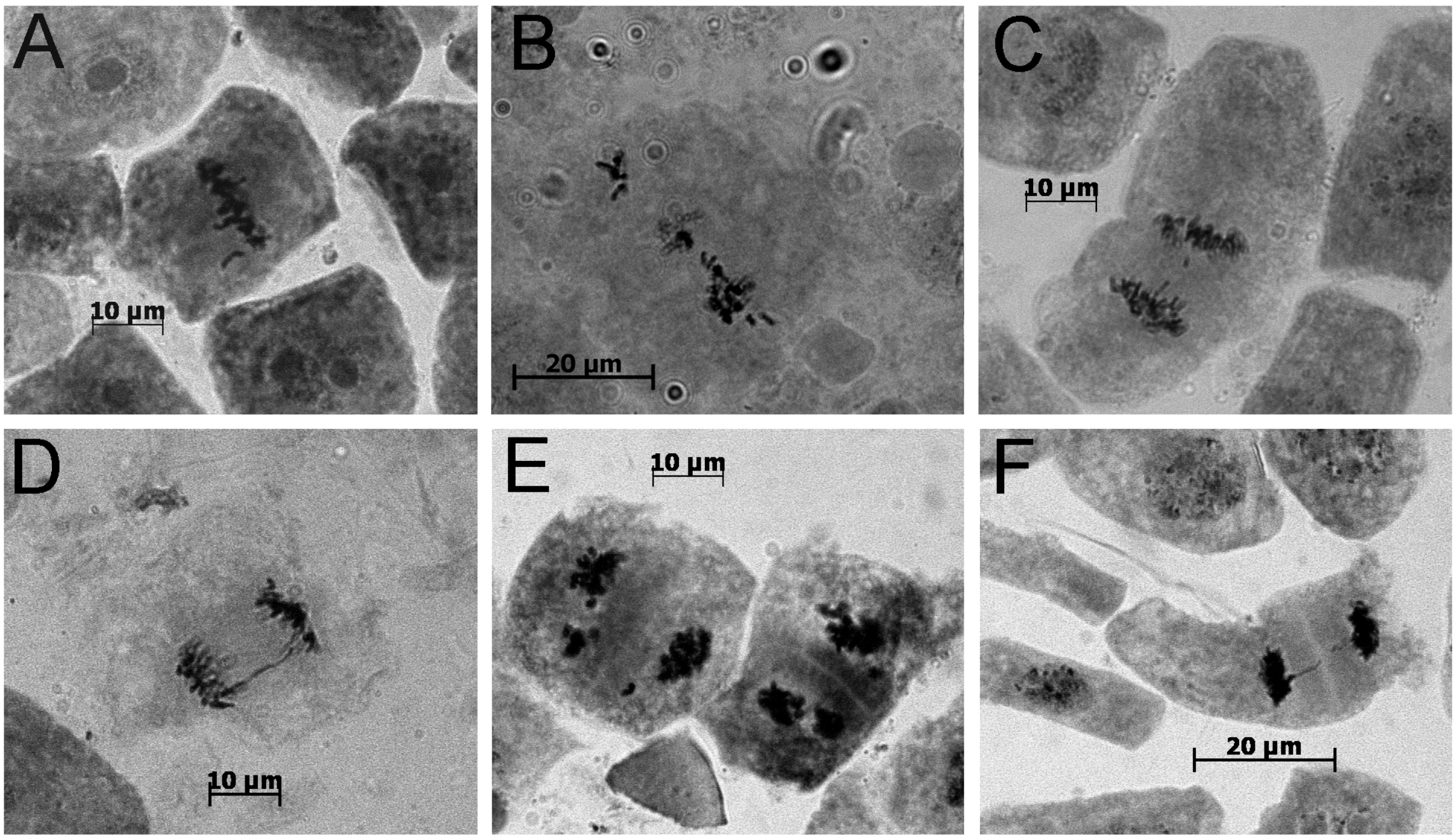

Fig. 4. Examples of abnormal mitoses in meristematic cells of seedlings. (A) Lagging of chromosomes in prometaphase. (B) Irregular grouping of chromosomes in metaphase. (C) Lagging of chromosomes in anaphase. (D) A bridge in anaphase. (E) Several groups of chromosomes in telophase. (F) A broken bridge in telophase.

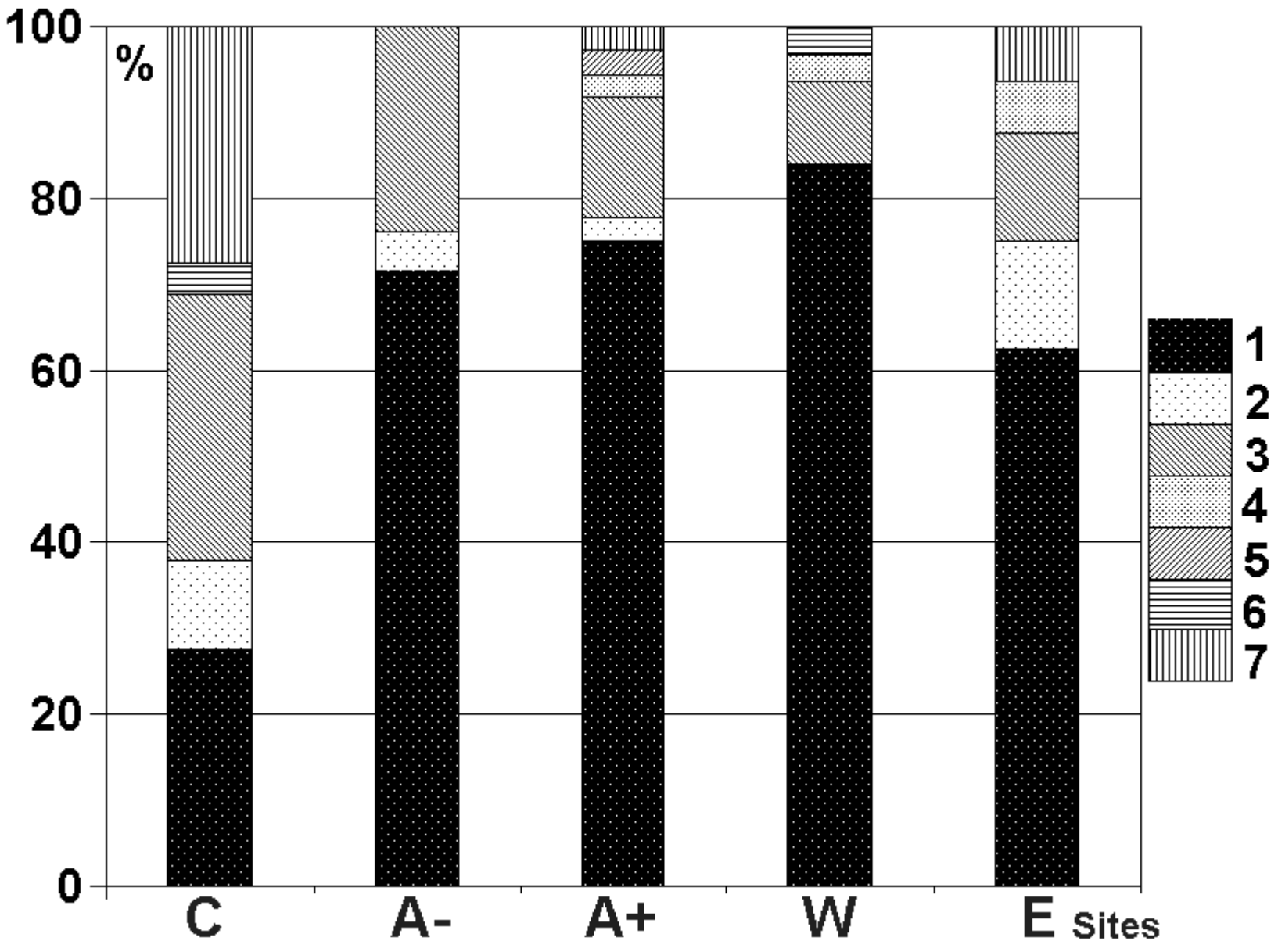

Fig. 5. The range of abnormal mitoses in meristematic cells of seedlings depending on the fruit sampling location. Lagging of chromosomes in prometaphase (1), irregular grouping of chromosomes in metaphase (2), lagging of chromosomes in anaphase (3), bridges in anaphase (4), chromosome ejections in anaphase (5), broken bridges in telophase (6), groups of chromosomes in telophase (7).

## 5. Discussion

The mitotic index reflects the division activity; its change may indicate an impact of unfavorable factors on the studied object. Responses of the apical meristem cells may be different depending on the nature of the stress and the character of the damage. A relatively weak level of stress factors can act as a catalyst for the process of cell division, but an increase of stress can suppress mitotic activity (Butorina, Kalaev, 2000; Vostrikova, Butorina, 2006).

The increase in the mitotic activity in meristematic cells from the W and E sites is due to the increasing proportion of cells at all mitotic phases, but the maximum increase was observed in prophase. This can occur because of a delay in the mitotic checkpoint G2/M. In order to overcome it, there is a need to complete DNA replication: the start of mitotic division is usually blocked when DNA is damaged or its replication is unfinished (Morgan, 2007).

The number of mitotic abnormalities in the later phases of mitosis is considerably reduced due to prophase and metaphase extension and amplification of repair processes of cellular systems. The increase in the meristematic activity is considered as a compensatory mechanism of the meristem to the increased cell death resulting from abnormal mitoses. The increase in the mitotic index due to the increase in the cell number in prophase is caused by chromosomal damage that prevents the transition of cells from one phase to other (Alberts et al., 2002; Vostrikova, Butorina, 2006).

The proportion of abnormal mitoses reflects the degree of DNA damage, and indicates the intensity of the mutation process. We observed small proportions of abnormal mitoses in samples from the control site in all mitotic phases. This is a result of spontaneous mutations usually associated with plant response to changes in weather conditions (Butorina et al., 2001; Kalaev, Butorina, 2006). However, the significant increase in the proportion of abnormal mitoses in samples from the A-, A+, and E sites is most likely connected with the influence of factors of the geological environment. This is because all these sites are located in the same microclimatic conditions. For the W site, the proportion of abnormal mitoses was slightly higher than that for the control site. However, it is necessary to consider the highest mitotic activity found for the W site that could led to elimination of most of the cells with mitotic abnormalities.

Samples from different sites differ by the range of abnormal mitoses. Mitotic abnormalities can be divided (Alov, 1972) into two groups: (1) those associated with chromosomal damages (i.e., lagging of chromosomes in prometaphase, anaphase, and telophase as well as bridges in telophase); and (2) those associated with disturbances of the mitotic spindle formation (i.e., irregular grouping of chromosomes in metaphase and telophase). Abnormalities of the first group were most common.

For all the test sites, the increase in the proportion of abnormal mitoses was observed in metaphase. Lagging of chromosomes in prometaphase was the predominant type of mitotic abnormalities; its minimum number was observed for the C control site. This abnormality may be an indicator of "fresh" chromosomal rearrangement, and is associated with chromosomal damage (Zosimovich, Kunakh, 1975). Sporadic irregular grouping of chromosomes in metaphase, found for the C, A+, and E sites, are connected with mitotic spindle damage. This can lead to an uneven distribution of chromosomes between daughter cells.

Chromosome laggings were most commonly observed in anaphase; they were recorded for all the sites. It is believed that lagging of chromosomes in anaphase is also associated with their damage (Zosimovich, Kunakh, 1975; Butorina et al., 2001).

Chromosome bridges observed in samples from the A+, W, and E sites can be caused by chromosomal rearrangements (Belousov et al., 2012). There is the second checkpoint between metaphase and anaphase; a critical indicator is the state of the mitotic spindle, and, therefore, the start of anaphase is blocked if spindle is damaged (Morgan, 2007). This is also indicated by the increase in the metaphase index.

An increase in the proportion of mitotic abnormalities in anaphase is considered as a sign of a decline of reparative abilities (Belousov et al., 2012). However, bridge occurrence may indicate an increase in reparative abilities of objects and their possible adaptation to a stress (Butorina et al., 2001; Kalaev, Butorina, 2006). Such an apparent contradiction is explained by the fact that bridges are considered a consequence of the association of chromosome fragments and adhesion of their telomeric fragments (Sunderland, 1973).

In a few cases we observed lagging and irregular grouping of chromosomes in telophase. Such mitotic abnormalities may be a result of multipolar mitosis or chromosome misdistribution in metaphase. For the W site, bridges and broken bridges in anaphase as well as the increase in the mitotic index (like for the E site), indicate the adaptive reaction of plants to conditions of growth in the local fault zone and the increase in their resilience.

Apparently, most cells with abnormal mitoses dies quickly and is not involved in the further development of the organism as indicated by the low proportions of mitotic abnormalities in telophase. There were minor differences between the sites by this index.

The sites are marked by the same temperature and water regimes; soils have similar granulometric content, pH values, and soil organic matter content; there is the similar plant species composition. Thus, differences in cytogenetic indices of samples from the test and control sites reflect the impact of factors of the geological environment.

## 6. Conclusions

The *Lonicera caerulea* subsp. *altaica* population located in the active fault zone is characterized by the heterogeneity of cytogenetic indices. The seed progeny from the test sites differ by the proportion and range of abnormal mitoses of seedling cells as well as the duration of mitotic phases.

The increase in the mitotic activity of meristematic cells from the sites in the local fault zone is connected with the occurrence of the prophase–metaphase block to prevent consequences of an increased cell death (as a result of abnormal mitoses in these phases) and to compensate their losses by a greater number of divisions. We observed the increase in the proportion of abnormal mitoses in samples from almost all the test sites, compared with the control site. This demonstrates the increase in the mutation activity in these sites. The range of abnormal mitoses of samples from all the test sites shows the increase in the proportion of abnormalities in metaphase, compared to the control site where they can be equally found in metaphase, anaphase, and telophase. The increase in the mitotic activity and bridge occurrence in anaphase and telophase are the adaptive reaction of meristematic cells of the apical meristem of seedlings to factors associated with active tectonic processes.

The results demonstrate the intensification of the evolutionary mutation process in the active fault zone under the action of factors of the geological environment. The results may help to clarify the role of environmental conditions within tectonically active zones in microevolutionary processes.

## Acknowledgements

The study was partially supported by the RFBR grant 15-37-50800. The authors are grateful to A.N. Dmitriev, A.Y. Gvozdarev, and A.I. Syso for the organization help in field surveys.